\documentclass[prl,aps,twocolumn,floats,showpacspsfig]{revtex4}
\usepackage{amssymb}

\usepackage{epsfig}

\newcommand{\be}{\begin{equation}}
\newcommand{\ee}{\end{equation}}
\newcommand{\bea}{\begin{eqnarray}}
\newcommand{\eea}{\end{eqnarray}}

\newcommand{\p}{\partial}
\newcommand{\s}{\sigma}

\newcommand{\la}{\langle}
\newcommand{\ra}{\rangle}
\newcommand{\rd}{\mbox{d}}
\newcommand{\ri}{\mbox{i}}

\begin{document}
\title{Crossed spin-1/2 Heisenberg chains as a quantum  impurity problem}

\author{ S. A. Reyes and A.  M. Tsvelik}
\affiliation{ Department of  Physics, Brookhaven National
Laboratory, Upton, NY 11973-5000, USA}

\affiliation{ Department of Physics and Astronomy, Stony Brook
University, Stony Brook, NY 11794-3800, USA}
\date{\today}

\begin{abstract}
Using equivalencies between different models we reduce the model of two spin-1/2 Heisenberg chains crossed at one point to the model of free fermions. The spin-spin correlation function is calculated by summing the perturbation series in the interchain interaction. The result reveals a power law decay with a nonuniversal exponent.

\end{abstract}

\maketitle

\sloppy

 Presence of impurities  in interacting systems causes nonlinear effects which may result in  a nontrivial scaling of thermodynamic quantities and correlation functions. Examples of impurity models discussed in the literature include various versions of the Kondo problem (a quantum spin in a noninteracting metallic host) and  the Kane-Fisher  or Boundary Sine-Gordon problem (a local static potential in a one-dimensional Luttinger liquid). They have numerous experimental applications in physics of diluted magnetic alloys (\cite{TsW83},\cite{Schlott}) and in such areas as interedge tunnelling in Quantum Hall effect (see, for example, \cite{Ludwig},\cite{Lutexp}). As a rule the impurity scattering in these models scales to strong coupling. The latter fixed point is rather simple in nature (fully screened spin in the Kondo problem, severed chain in the Kane-Fisher one). The exclusion is the fixed point in the underscreened  Kondo problem where the fixed point occurs at  an intermediate coupling first predicted in \cite{Nozieres79}. This fixed point is characterized by non-trivial universal indices.

 In this paper we would like to call the attention to the situation when the operator describing a scattering on the impurity is exactly marginal (that is its scaling dimension is equal to the dimension of space-time, which in the present case is 2). Since such interaction does not flow under renormalization, the results for the correlation functions are bound to depend on the bare coupling constant. In particular,  such situation exists  when  the underlying impurity problem is equivalent to  a problem of noninteracting fermions scattering on a scalar potential. An interesting  situation emerges when the fermion operators and observables 
  are mutually nonlocal. In that case calculation of  correlation functions of the physical fields still constitutes a nontrivial problem resulting in nonuniversal scaling dimensions.


 One experimentally relevant  realization of the Marginal Quantum Impurity problem is provided by the model  of two spin-1/2 Heisenberg chains interacting
at a single point by the exchange interaction:
\bea
 && H =  \sum_{n=-\infty}^{+\infty}
J\left[{\bf S}_{1}(n)\cdot{\bf S}_{1}(n+1) + {\bf S}_{2}(n)\cdot{\bf S}_{2}(n+1)\right] \nonumber\\
&& + J_{\bot}{\bf S}_{1}(0)\cdot{\bf S}_{2}(0)
\eea
where $|J_{\bot}| << J$.
 This model can be treated as a particular version of the spin ladder problem.   To describe its low energy properties one can employ the technique developed  in \cite{ShNerTs96} (see also \cite{book} for a more detailed description) and rewrite the Hamiltonian as the model of four species of massless  real (Majorana) fermions coupled at point $x =0$
\bea
H = \int \rd x \sum_{a=0}^3\left[\frac{\ri v}{2}(-R_a\p_x R_a + L_a\p_x L_a) + \ri g_a\delta(x)(R_aL_a)\right] \label{model1}
\eea
where $g_i = J_{\bot} a_0 (i=1,2,3); g_0 = - 3g_1$ and $v= \pi Ja_0/2$ is the spinon velocity. The fermion operators satisfy the standard anticommutation relations 
\bea
&& \{R_a(x),R_b(y)\} = \{L_a(x),L_b(y)\} = \delta_{ab}\delta(x-y)\nonumber\\
&& \{R_a(x),L_b(y)\} = 0
\eea
and are real, that is $R^+ = R, L^+ =L$. The ratio $g_0/g_a$ can be changed by introduction of the four-spin interaction \cite{NerTs97}. Thus  the model of interacting spins is reduced to the model  of non-interacting fermions. This representation respects the original symmetry of the problem: the fermions $a=1,2,3$ transform as an SU(2) triplet and the 0-th fermion is an SU(2) singlet. Fermionization of one-dimensional spin models has a long history going back to the work by Jordan and Wigner \cite{wigner}. It is well known that a single spin-1/2 Heisenberg chain can be represented as a model of fermions which interaction depends on the anisotropy. At the isotropic point this interaction is quite strong. Therefore it is interesting to note that though a single isotropic spin-1/2 Heisenberg chain cannot be described as a model of noninteracting fermions, the two chain model  can. Naturally, the spin operators of the original Heisenberg  chains are nonlocal with respect to the Majorana fermions (the relationship between them resembles the one given by the Jordan-Wigner transformation). For that reason the problem of correlation functions still remains nontrivial. To calculate the spin correlators we will employ an alternative  representation of model (\ref{model1}), namely, in the form of four quantum Ising models:
\bea
H = \sum_{a=0}^3 H_{Is}^a, ~~ H_{Is}^a = H_{crit}^{a} + g_a\epsilon_a(x=0) \label{model2}
\eea
where $H_{crit}$ is the Hamiltonian of the critical Ising model and $\epsilon(x,\tau)$ is the energy density field. The quantum Ising model is described by the Hamiltonian 
\bea
H = -J\sum_n(\sigma_n^z\sigma_{n+1}^z + h\sigma_n^x) \label{Ising}
\eea
The Jordan-Wigner transformation brings it to the fermionic form.  The order parameter field $\s(x)$ of the Ising model is the continuum limit of $\sigma^z_n$, the energy density field is the continuum limit of $\sigma^x$. At $h <1$  field  $\s$ has a nonzero vacuum average $\la\s\ra \neq 0$. Hamiltonian  (\ref{Ising}) can be rewritten in the dual form
\bea
H = -J\sum_n(h\mu_{n-1/2}^z\mu_{n+1/2}^z + \mu_{n+1/2}^x) \label{Ising1}
\eea
where the operators 
\bea
\mu^z_{n +1/2} = \prod_{j\leq n}\sigma_j^x, ~~ \mu_{n+1/2}^x = \sigma_n^z\sigma_{n+1}^z
\eea
obey the same commutation relations as the Pauli matrices $\s^z,\s^x$. The so-called disorder parameter field $\mu(x)$ is defined as the continuum limit of the operator $\mu^z_{n+1/2}$. It is clear that $\la\mu\ra \neq 0$ at $h >1$.  At $h=1$  the models (\ref{Ising}) coincides with its dual (\ref{Ising1}). Since $\s$ and $\mu$ cannot have nonzero ground state expectation values simulataneously, at $h=1$ their averages vanish and the model is  quantum critical. At this point the Majorana fermion becomes massless. Thus model (\ref{Ising}) with $h=1$ is equivalent to the model of one species of massless Majorana fermions.

The advantage of the Ising model  representation is that the original spin fields of the Heisenberg models can be written as 
\bea
&& {\bf S}_{1}(j) + {\bf S}_{2}(j)= \frac{\ri}{2}\left\{[{\bf R}\times{\bf R}] + [{\bf L}\times{\bf L}]\right\} + (-1)^j{\bf n}_+(x)\nonumber\\
&& {\bf S}_{1}(j) - {\bf S}_{2}(j)= \frac{\ri}{2}\left\{R_0{\bf R}
+  L_0{\bf L}\right\} + (-1)^j{\bf n}_-(x) \eea 
where the most
relevant parts of the spin operators given by the staggered
magnetizations ${\bf n}_{\pm}$ are expressed as local combinations of the order and disorder parameters of the Ising models \cite{ShNerTs96}:
 \bea
n^x_{+} = \s_1\mu_2\s_3\mu_0, ~~ n^y_+ = \mu_1\s_2\s_3\mu_0, ~~
n^z_+ = \s_1\s_2\mu_3\mu_0 \label{n}
\eea 
In the expression for ${\bf n}_-$
one has to interchange $\s$ and $\mu$. Correlation functions of the Ising model fields and their properties are well known and we are going to use this knowledge to calculate the correlators of the perturbed model (\ref{model1}).   Using Eqs.(\ref{n})  and
taking into account that the change of sign of the coupling
constant $g \rightarrow -g$ is equivalent to the substitution $\mu
\rightarrow \s, \s \rightarrow \mu$, it is easy to relate  the
desired spin correlators to the correlation functions of the perturbed
Ising model: \bea
&& \langle n_{\alpha}^{a}(\tau_1,x_1) n_{\beta}^{a}(\tau_2,x_2) \rangle = \label{G}\\
&& G_{\sigma,g}^{2}(\tau_{12};x_1,x_2)G_{\mu,g}(\tau_{12};x_1,x_2)G_{\sigma,3g}(\tau_{12};x_1,x_2)
+ \nonumber\\
&& (2\delta_{\alpha\beta}-1)G_{\mu,g}^{2}(\tau_{12};x_1,x_2)G_{\sigma,g}(\tau_{12};x_1,x_2)\times \nonumber\\
&& G_{\mu,3g}(\tau_{12},x_1,x_2) \nonumber \eea 

where,
\bea
G_{\mu,g}(\tau;x_1,x_2) \equiv \la\langle\mu(\tau,x_1)\mu(0,x_2)\rangle\ra \\
G_{\sigma,g}(\tau;x_1,x_2) \equiv \la\langle\sigma(\tau,x_1)\sigma(0,x_2)\rangle\ra
\eea

and $\alpha$,$\beta$ label the chain to which the operator corresponds (1 or
2). Notice that the correlators remain translationally invariant
only in time direction. To simplify the calculations  we will consider the
above correlation functions only at $x_{1,2} =0$. To obtain  these correlators we sum the leading logarithms in the  perturbation expansion in  small $g_a$.

Namely, the Ising order parameter field correlator,
${\langle\la\sigma(\tau_a)\sigma(\tau_b)\rangle\ra}$ (we omit the space coordinate $x$ assuming $x =0$) can be obtained by
calculating the following series,
\bea \label{sum_c}
 \la\la\sigma(\tau_a)\sigma(\tau_b)\ra\ra_g =
\sum_{n=0}^{+\infty}\frac{g^n}{n!} C_n
\eea
where,
\bea
C_n \equiv \int \rd \tau_1 \ldots \rd \tau_n
\langle\langle\sigma(\tau_a)\sigma(\tau_b)\varepsilon(\tau_1)\ldots\varepsilon(\tau_n)\rangle\rangle_0
\eea
and  ${\langle\langle \rangle\rangle_0}$ denotes the irreducible
correlator in the unperturbed system and,
 $ \varepsilon \equiv \ri RL(x=0)$
is the Ising model energy density at the impurity point.

Only the largest divergent terms will be kept at each order in
$g$. The irreducible correlators under consideration will have its
largest divergencies in the regions where  each of the ${\tau_{i}
\;'s}$ approach either ${\tau_a}$ or ${\tau_b}$ corresponding to the  fusion of $\epsilon$ and $\s$ operators. Divergencies corresponding to the  fusion of $\epsilon$ operators are not present in the irreducible correlation functions being  cancelled by the corresponding  divergencies in the partition function. To calculate the leading logarithms we take advantage of  the Operator Product Expansion (OPE) for the critical Ising model \cite{ope}:

\bea \label{ope_sig}
 \varepsilon(\tau_i)\sigma(\tau_{a,b}) = \frac{1}{2\vert\tau_{a,b} -
\tau_i\vert}\sigma(\tau_{a,b}) + ...
\eea
where the dots stand for less relevant terms.
The most divergent part of $C_n$ is given by
\bea
 C_n \approx \left[ 2 \ln \left(\frac{\vert
\tau_{ab} \vert}{\tau_0}\right) \right]^n
 \eea
where $\tau_0$ is an ultraviolet cutoff. The factor $2^n$ comes
from the number of regions with divergent integrand that exist
(that is to say, the number of ways the $n$ different $\tau_i$
variables can approach either ${\tau_a}$ or ${\tau_b}$) and the
logarithm comes from integrating over this regions.

Now, replacing this result into (\ref{sum_c}) is easy to see that,
\bea \label{s}
 &&G_{\sigma,g} = \langle\sigma(\tau_a)\sigma(\tau_b)\rangle_0
\left(\frac{\tau_0}{\vert \tau_{ab} \vert}\right)^{-2g_1} \\
&& = \frac{1}{\vert \tau_{ab}
\vert^{\frac{1}{4}}}\left(\frac{\tau_0}{\vert \tau_{ab}
\vert}\right)^{-2g_1} \nonumber
\eea
where
$\langle\sigma(\tau_a)\sigma(\tau_b)\rangle_0 \sim |\tau_{ab}|^{-1/4}$ is  the
correlator of the unperturbed system.

Similar considerations are valid for the perturbation series for
the disorder parameter, $\langle\mu(\tau_a)\mu(\tau_b)\rangle$.
The only difference is that the OPE contains minus sign \cite{ope}:
\bea \label{ope_mu}
\varepsilon(\tau_i)\mu(\tau_{a,b}) = -\frac{1}{2\vert\tau_{a,b} -
\tau_i\vert}\mu(\tau_{a,b})  + ...\eea
Then the same steps
that lead to (\ref{s}), now lead to:
 \bea \label{mu}
&& G_{\mu, g} = \langle\la\mu(\tau_a)\mu(\tau_b)\ra\rangle_0 \left(\frac{\tau_0}{\vert
\tau_{ab} \vert}\right)^{2g} \\ 
&& = \frac{1}{\vert \tau_{ab}
\vert^{\frac{1}{4}}}\left(\frac{\tau_0}{\vert \tau_{ab}
\vert}\right)^{2g} \nonumber
\eea
where $g = g_1$ for $\mu_a$ (a=1,2,3) operator and $g_0$ for the $\mu_0$ operator. 

Substituting (\ref{s},\ref{mu}) into (\ref{G}) we obtain
\bea
 && \langle {\bf n}_{\alpha}(\tau_a) \cdot {\bf n}_{\beta}(\tau_b) \rangle = \\
&&\frac{3}{\vert \tau_{ab}
\vert}\left[\left(\frac{\tau_0}{\vert \tau_{ab}\vert}\right)^{-8\tilde g}
+
(2\delta_{\alpha\beta}-1)\left(\frac{\tau_0}{\vert\tau_{ab}\vert}\right)^{8\tilde g}\right]\nonumber
 \eea
where $\tilde g = (g_1 - g_0)/4$. 
 The nonuniversal power law behavior of the spin-spin correlation functions is reflected in the power law behavior of the local  magnetic susceptibility:
\bea
\chi(x=0) \sim (T/T_0)^{8\tilde g} + (T/T_0)^{-8\tilde g}, ~~ T_0 \sim J \label{chi}
\eea
As one may expect, the susceptibility does not depend  on the sign of the interchain interaction. Since the perturbing impurity operator is exactly marginal, it does not generate any nontrivial corrections to the specific heat. The impurity magnetic susceptibility diverges with a nonuniversal index (\ref{chi}).

 Thus we have found a nontrivial  example of the impurity problem where strong correlations in the bulk generate nonuniversal scaling dimensions. This model may be a member of a class of models which  Hamiltonians can be written as a scattering problem of free particles, whose creation and annihilation operators and  the observables are mutually nonlocal. 

This research was supported by US DOE under contract numbers
DE-AC02 -98 CH 10886. S. A. R. is grateful to  Robert Konik for many useful discussions.

\end{document}